\documentclass[aps,11pt,nofootinbib]{revtex4-1}
 \pdfoutput=1
\usepackage{graphicx}
\usepackage[margin=1in]{geometry}
\usepackage{amsmath}
\usepackage{amssymb}
\usepackage{color}
\usepackage[utf8]{inputenc}
\usepackage[normalem]{ulem}
\usepackage{siunitx}
\usepackage{booktabs}
\usepackage{tabu}

\makeatletter
\renewcommand\@make@capt@title[2]{%
  \@ifx@empty\float@link{\@firstofone}{\expandafter\href\expandafter{\float@link}}%
   {\textbf{#1}}\@caption@fignum@sep#2\quad
}%
\makeatother

\begin{document}

\author{Sam Young}
\email{young@mail.lorentz.leidenuniv.nl}

\affiliation{Instituut-Lorentz for Theoretical Physics, Leiden University, 2333 CA Leiden, The Netherlands}
\affiliation{Max Planck Institute for Astrophysics, Karl-Schwarzschild-Strasse 1, 85748 Garching bei Muenchen, Germany}

\date{\today}

\title{Peaks and primordial black holes: the effect of non-Gaussianity}

\begin{abstract}
In light of recent developments in the field, we re-evaluate the effect of local-type non-Gaussianity on the primordial black hole (PBH) abundance (and consequently, upon constraints on the primordial power spectrum arising from PBHs). We apply peaks theory to the full, non-linear compaction, finding that, whilst the effect of non-Gaussianity is qualitatively similar to previous findings, the effect is much less significant.  It is found the non-Gaussianity parameters $f_\mathrm{NL}^\mathrm{local}$ and $g_\mathrm{NL}^\mathrm{local}$ typically need to be approximately 1 or 2 orders of magntiude larger respectively to have a similar to that previously found. The effect will be to weaken the dependance of PBH constraints on the primordial power spectrum on the non-Gaussianity parameters, as well as to dramatically weaken constraints on the non-Gaussianity parameters (and/or PBH abundance) arising from the non-observation of dark matter isocurvature modes. We also consider the correlation between the curvature perturbation $\zeta$ and the compaction $C$, finding that, whilst PBHs may form at rare peaks in $C$ these do not necessarily correspond to rare peaks in $\zeta$ - casting some doubt on many of the existing calculations of the PBH abundance.
\end{abstract}

\maketitle

%
\section{Introduction}
%

Primordial black holes (PBHs) are black holes which may have formed in the early universe.  The possible existence of PBHs was initially considered by Novikov and Zel'Dovic \cite{1967SvA....10..602Z}, followed shortly by work from Hawking and Carr \cite{Hawking:1971ei,Carr:1974nx,Carr:1975}. PBHs are of great interest cosmologically, because, in addition to being a viable dark matter candidate \cite{Carr:2009jm,Carr:2020xqk,Carr:2020gox}, PBHs provide unique constraints on the primordial power spectrum, and have been proposed to be the source of the gravitational waves from merging black holes observed by LIGO-Virgo \cite{Clesse:2017bsw,DeLuca:2019buf,Mirbabayi:2019uph,Postnov:2019tmw,Fernandez:2019kyb,He:2019cdb,LIGOScientific:2018jsj,LIGOScientific:2018mvr,DeLuca:2020bjf}. Numerous mechanisms have been proposed for their formation (for example, from cosmic strings \cite{Hawking:1987bn} or bubble collisions \cite{Hawking:1982ga}, see \cite{Green:2014faa} for a review) although we will here focus on PBHs which form from the gravitational collapse of large density perturbations.  

Carr's initial work \cite{Carr:1974nx} calculated that if the density contrast is above some threshold value $\delta_c$ then that region will collapse to form a PBH when it enters the horizon.  An order of magnitude estimate was performed, finding that the density contrast should be greater than the equation of state in order for a PBH to form, $\delta_c=\omega$ (with $\omega=1/3$ during radiation domination).  However, there has since been an extensive amount of work to determine the collapse threshold, settling on a slightly larger value $\delta_c\simeq 0.5$ \cite{Musco:2004ak,Musco:2008hv,Musco:2012au,Musco:2018rwt,Harada:2015ewt,Harada:2015yda,Nakama:2013ica,Nakama:2014fra,Shibata:1999zs,Niemeyer:1999ak,Polnarev:2006aa,Escriva:2019phb,Escriva:2020tak}. This formation threshold is orders of magnitude larger than perturbations seen in the CMB, and so in order for a significant number of PBHs to form, the power spectrum on scales which form PBHs must also be orders of magnitude larger than observed on cosmological scales. There are many models which do make this prediction (for example, \cite{Drees:2011hb,Bugaev:2013fya,Ozsoy:2018flq,GarciaBellido:1996qt,Lyth:2012yp,Bugaev:2011wy,Ballesteros:2018wlw}, amongst many others).

Researchers building models of inflation typically make predictions for the power spectrum (and higher-order correlation functions) in terms of the curvature perturbation $\zeta$, which appears as a perturbative quantity in the FLRW metric in the comoving uniform-density gauge as
\begin{equation}
\mathrm{d}s^2 = -\mathrm{d}t^2 + a^2(t)\exp\left( 2 \zeta \right)\mathrm{d}\mathbf{X}^2,
\label{eqn:metric}
\end{equation}
where $\zeta$ can be seen to be an effective rescaling of the physical coordinate $\mathbf{X}$. When we want to predict the abundance of PBHs given a particular model, or else constrain a particular model using constraints on PBHs, it is desirable to be able to relate the power spectrum and non-Gaussianity of $\zeta$ to the PBH abundance (see \cite{Carr:2020gox,Gow:2020bzo} for a recent compilation of constraints on the PBH abundance and consequent constraints on the power spectrum). 

It has been shown that primordial non-Gaussianity can have a strong impact on the abundance of PBHs \cite{Bullock:1996at,Ivanov:1997ia,Byrnes:2012yx,Shandera:2012ke,Young:2013oia,Young:2015cyn,Franciolini:2018vbk,Yoo:2020dkz,Yoo:2019pma,Atal:2019cdz,Atal:2018neu,Riccardi:2021rlf,Kitajima:2021fpq}, as well as the mass function and initial clustering \cite{Young:2019gfc}.  It has also been argued recently that, in models which predict large numbers of PBHs, there is expected to be a large amount of non-Gaussianity \cite{Figueroa:2020jkf,Biagetti:2021eep}. It is therefore important to correctly account for the effect of non-Gaussianity when deriving constraints of the power spectrum from PBHs. In this paper, we will reconsider the effect of non-Gaussianity on the PBH abundance, accounting for recent developments in the field.

The layout of the paper is as follows: in section \ref{sec:criteria} we will discuss the formation criteria for PBHs in the context of local-type non-Gaussianity, in section \ref{sec:variables} we will discuss the technical details of the calculation and introduce variables, section \ref{sec:NGabundance} calculates the effect of local-type non-Gaussianity on the PBH abundance, a comparison is made to previous literature in section \ref{sec:literature}, and finally section \ref{sec:conclusions} contains the conclusions reached in the paper.

%
\section{Formation criteria}
\label{sec:criteria}
%

The most suitable parameter to use to determine whether a perturbation will collapse to form a PBH is the compaction function $C$ \cite{Young:2014ana,Musco:2018rwt,Young:2019osy}, which is defined as
\begin{equation}
C(\mathbf{x},r) \equiv 2 \frac{\delta M(\mathbf{x},r,t)}{R(r,t)},
\end{equation}
where $\delta M(\mathbf{x},r,t)=M(\mathbf{x},r,t)-M_b(\mathbf{x},r,t)$ is the mass excess within a sphere areal radius $R(\mathbf{x},r,t)=a(t)\exp(\zeta(\mathbf{x}))r$ centred on spatial coordinate $\mathbf{x}$; $M(\mathbf{x},r,t)$ is the Misner-Sharp mass and the subscript $b$ denotes the background value in a region of unperturbed space. The compaction is closely related to the density constrast, and is used to describe the amplitude of density perturbations with a time-independent parameterisation (whilst the individual components of the compaction are time-dependent,  the overall function is not).  The compaction is discussed in more detail in appendix \ref{app:compaction}, and the benefits of using such a parameter are also described in detail in reference \cite{Young:2019osy}.

If the compaction is above some critical value $C_\mathrm{th}$ in a region whilst it is super-horizon,  a PBH will form once the regions re-enters the horizon,  with a mass given by
\begin{equation}
M_\mathrm{PBH}\left( C \right) = K M_\mathrm{H}\left( C - C_\mathrm{th} \right)^\gamma,
\label{eqn:mass}
\end{equation}
where $M_H$ is the horizon mass or the unperturbed background at the time when the horizon scale is equal to the smoothing scale used (in comoving units), and we take the values $K=4$, $C_{th}=0.5$ and $\gamma=0.36$ \cite{Young:2019yug}.

There are several different approaches which can be used to calculate the PBH abundance. The simplest,  a Press-Schechter-type approach (also referred as the threshold statistics approach),  simply states that PBHs form in regions where the compaction at a given point is above the threshold.  Peaks theory provides a more accurate calculation by adding the condition that PBHs will form at locations where the compaction function is at a maximum (i.e. where there is a peak in the compaction).  Recent developments have also included a condition to determine the scale of peaks in the compaction - since the mass of a PBH depends non-trivially on both the scale and amplitude of a perturbation \cite{Young:2020xmk,Germani:2019zez}.  The peak constraint $c_\mathrm{pk}$ describes these criteria and gives the number of peaks (either 1 or 0) of height $\bar{C}$ and scale $r$ in the infinitesimal volume $\mathrm{d}^3x\mathrm{d}r$,
\begin{equation}
c_\mathrm{pk} = \delta_D\left(C-\bar{C}\right) \delta_D^{(3)}\left( \bar{\nabla}_i C \right)  \Theta_H\left( \lambda_3 \right) \delta_D\left( \frac{\mathrm{d}C}{\mathrm{d}r}  \right)\Theta_H\left(- \frac{\mathrm{d}^2 C}{\mathrm{d}r^2}  \right),
\label{eqn:peakConstraint}
\end{equation}
where $\delta_D^{(n)}$ is the $n$-dimensional Dirac-delta function, $\Theta_H$ is the Heaviside step function,  and $\lambda_3$ is the smallest eigenvalue of $\nabla_i\nabla_jC$.  For sufficiently narrow spectrum, the critieria to determine the scale of the perturbation is unnecessary, essentially since all perturbations have the same characteristic scale,  and the PBH abundance can be calculated by considering only this single scale \cite{Young:2020xmk,Germani:2019zez}.

If the probability density function (PDF) of the compaction $C$ and its derivatives are known,  the number of peaks of given height and scale can be calculated,  and from there the PBH abundance.  In general, this is problematic,  because there is no analytic expression for the compaction in terms of $\zeta$. However,  in the high-peak limit, the simplifying assumption is usually made that peaks are spherically symmetric, which is considered a suitable approximation since rare peaks that form PBHs are expected to be spherically symmetric \cite{Bardeen:1985tr}.  The validity of this assumption is discussed in appendix \ref{sec:validity}.  Under the assumption of spherical symmetry, the compaction can be expressed in terms of the linear component $C_1(\mathbf{x},r)$ \cite{Harada:2015yda} (a brief derivation of this is found in appendix \ref{app:compaction}):
\begin{align}
C(\mathbf{x},r) &=- \frac{4}{3}r\zeta'(r)\left( 1+\frac{1}{2}r\zeta'(r) \right),\\
&= C_1(\mathbf{x},r)-\frac{3}{8}C_1(\mathbf{x},r)^2.
\label{eqn:quadratic}
\end{align}

There is a maximum value for the compaction, $C_\mathrm{max}=2/3$ which occurs when $C_1=4/3$.  Perturbations with $C_1<4/3$ are referred to as type I perturbations and, if above the threshold value, can form PBHs with a mass dependent on both the scale and amplitude of the perturbation.  Perturbations with $C_1>4/3$ are referred to as type II perturbations (whereby the areal radius does not increase monotonically with the radial coordinate $r$). It was previously thought that such perturbations did not form PBHs (instead forming separate universes), but reference \cite{Kopp:2010sh} showed rather that type II perturbations always lead to the formation of PBHs. However,  it is not possible to simulate the formation of such PBHs using the density, and as a result, the formation and resultant mass of such PBHs is not well understood. In addition, the abundance of type II perturbations is exponentially suppressed compared to type I perturbations, and so we will neglect type II perturbations for the remainder of this paper. With this in mind, inverting equation \eqref{eqn:quadratic} and keeping only the relevant solution gives 
\begin{equation}
C_1 = \frac{4}{3}\left( 1 - \frac{         \sqrt{2-3  C}      }{\sqrt{2}} \right).
\label{eqn:C1ofC}
\end{equation}

\subsection{Effect of non-Gaussianity}

In this section,  we will consider the effect of local-type non-Gaussianity on the compaction function. The effect of non-Gaussianity on the profile shape and threshold for collapse was studied in reference \cite{Kehagias:2019eil} finding that non-Gaussianity has a small impact on the formation threshold, and we will thus neglect these effects for the remainder of this paper. We can express the effect of local-type non-Gaussianity by making writing the curvature perturbation $\zeta$ as a series in terms of Gaussian variable $\zeta_G$
\begin{equation}
\zeta = \zeta_G+f\left( \zeta_G^2-\langle \zeta_G^2 \rangle \right)+g\zeta_G^3+\cdots,
\label{eqn:localNG}
\end{equation}
where $f=3f_\mathrm{NL}^\mathrm{local}/5$ and $g=9g_\mathrm{NL}^\mathrm{local}/25$ describe the level of non-Gaussianity.  Higher-order terms are neglected here,  but were studied in reference \cite{Young:2013oia}, finding the effects of odd- or even-order terms to be qualitatively similar to the quadratic and cubic terms respectively (although our results here will suggest that they have a much smaller effect). The  $\langle\zeta_G^2 \rangle$ term is included such that the expecation value remains zero, $\langle\zeta\rangle=0$.  The quadratic term introduces skewness to the distribution, whilst the cubic term affects the kurtosis.  It is worth noting that an expansion such as this may not be valid when studying PBHs, since higher-order terms can have a large impact and may not be neglectable \cite{Young:2013oia,Atal:2019erb,DeLuca:2022rfz} - although we will see later in the calculation presented here that the effect of higher order terms will be suppressed compared to previous calculations. 

Peaks in $\zeta_G$ typically correspond to peaks in $\zeta$, but can be troughs depending on the amplitude of the peak and the values of $f$ and $g$ (this will be discussed in more detail in section \ref{sec:NGabundance}).  In addition, in the high-peak limit which will be relevant for studying PBH formation, we can continue to make the assumption that relevant peaks are spherically symmetric - in which case peaks in one variable will correspond to peaks (or troughs) in any other variable we will consider, such as the density or compaction.

Neglecting higher-order terms, we can now express the linear component of the compaction as
\begin{equation}
C_1(\mathbf{x},r)=-\frac{4}{3}r\zeta'(r) =-\frac{4}{3} r\zeta_G'(r)\left(1+2 f\zeta_G(r)+3g \zeta_G(r)^2\right),
\label{eqn:c1}
\end{equation}
which can be substituted into equation \eqref{eqn:quadratic} to give the full expression for the compaction.

In principle, the number density of peaks of given scale and amplitude (which then gives the abundance of PBHs of given mass), can be calculated by numerically integrating the peak constraint, equation \eqref{eqn:peakConstraint}, over the probability density function (PDF) of the relevant variables and their first and second derivatives (see i.e.  section III of \cite{Young:2020xmk} for more information).  In the following sections, we will formulate a simple procedure to calculate the abundance of PBHs by taking the high-peak limit.

%
\section{Variables,  correlation factors, and probability density functions}
\label{sec:variables}
%

In the context of local-type non-Gaussianity, the compaction depends on the terms $-\frac{4}{3}r\zeta_G'(r)$ and $\zeta_G(r)$,  and the PDF of the compaction $P_C(C)$ can therefore be expressed by making use of the (Gaussian) PDFs of these terms.

The term $-\frac{4}{3}r\zeta_G'(r)$ is the expression that one would obtain by convolving the linear expression for the density contrast, $\delta_l=-4 \nabla^2\zeta_G/9$, with a top-hat smoothing function $W(\mathbf{x},r)$ \cite{Young:2019yug} at the centre of a spherically symmetric peak.  We will refer to this quantity as the Gaussian component of the compaction, $C_G$:
\begin{equation}
C_G(\mathbf{x},r) = -\frac{4}{9}r^2\int\mathrm{d}^3\mathbf{y} \nabla^2\zeta_G(\mathbf{y})W\left( \mathbf{x}-\mathbf{y} ,r \right) = -\frac{4}{3}r\zeta_G'(r),
\end{equation}
where $\mathbf{x}$ is the centre of a peak (corresponding to $r=0$), and spherical symmetry has been assumed in the second equality.  The smoothing function is given by
\begin{equation}
W(\mathbf{x},r) = \frac{3}{4\pi r^3}\Theta_\mathrm{H}\left( r-x \right),
\end{equation}
where $\Theta_\mathrm{H}(x)$ is the Heaviside-step function.  The Fourier transform of this window function is
\begin{equation}
\tilde{W}(k,r) = 3\frac{\sin(kr)-kr\cos(kr)}{(kr)^3}.
\label{eqn:FourierTH}
\end{equation}

Similarly,  since the $\zeta_G(r)$ term originates from the surface term of an integral over a sphere of radius $r$,  it can be expressed as a smoothing of the curvature perturbation with a spherical-shell function. We will refer to this term as $\zeta_r$:
\begin{equation}
\zeta_r(\mathbf{x}) = \int\mathrm{d}^3\mathbf{y} \zeta_G(\mathbf{y})W_s\left( \mathbf{x}-\mathbf{y} ,r \right) = \zeta_G(r),
\end{equation}
where $\mathbf{x}$ is the centre of a peak (corresponding to $r=0$), and spherical symmetry has been assumed in the second equality. The spherical-shell window function $W_s$ is given by
\begin{equation}
W_s(\mathbf{x},r) =\frac{1}{4\pi r^2} \delta_D\left(x-r \right),
\end{equation}
with Fourier transform given by
\begin{equation}
\tilde{W}_s(k,r) = \frac{\sin(kr)}{kr}.
\end{equation}

For ease of reference, we will now define some of variables which will be used throughout the remainder of the paper. Firstly, we define the following integrals over the Gaussian component of the power spectrum, which will be needed for the covariance matrix and calculation of the PBH abundance later:
\begin{align}
\begin{split}
\sigma_{\zeta}^2 &= \int\limits_0^\infty \frac{\mathrm{d}k}{k}\mathcal{P}_{\zeta_G},\\
\sigma_c^2 &= \frac{16}{81}\int\limits_0^\infty \frac{\mathrm{d}k}{k}(kr)^4  \tilde{W}^2(k,r)\mathcal{P}_{\zeta_G},\\
\sigma_n^2 &= \frac{16}{81}\int\limits_0^\infty \frac{\mathrm{d}k}{k}(kr)^4  \tilde{W}^2(k,r)k^{2n}\mathcal{P}_{\zeta_G},\\
\sigma_r^2 &= \int\limits_0^\infty \frac{\mathrm{d}k}{k} \tilde{W}_s^2(k,r) \mathcal{P}_{\zeta_G},\\
\sigma_{cr}^2 &= \frac{4}{9}\int\limits_0^\infty \frac{\mathrm{d}k}{k}(kr)^2 \tilde{W}(k,r)\tilde{W}_s(kr) \mathcal{P}_{\zeta_G},\\
\sigma_{c\zeta}^2 &= \frac{4}{9}\int\limits_0^\infty \frac{\mathrm{d}k}{k}(kr)^2 \tilde{W}(k,r) \mathcal{P}_{\zeta_G}.
\end{split}
\label{eqn:moments}
\end{align}
We will also define the following variables, with unit variance:
\begin{equation}
\nu_c = \frac{C_G}{\sigma_c}, \nu_r=\frac{\zeta_r}{\sigma_r}, z = \frac{\nu_r-\gamma_{cr}\nu_c}{\sqrt{1-\gamma_{cr}^2}},
\end{equation}
where 
\begin{equation}
\gamma_{cr}=\frac{\sigma_{cr}^2}{\sigma_c\sigma_r},
\end{equation}
is the correlation coefficient of $\nu_c$ and $\nu_r$. The reason for introducing the variable $z$ is to diagonalise the 2-variate Gaussian appearing in the next section.

\subsection{The 2-variate Gaussian probability density function}

The PDF of $C_G$ and $\zeta_r$ can be described with a 2-variate Gaussian
\begin{equation}
\mathcal{N}\left( \mathbf{Y} \right) = (2\pi)^{-1/2}\det\left(\mathbf{\Sigma}\right)^{-1/2} \exp\left( -\frac{1}{2}\mathbf{Y}^T \mathbf{\Sigma}^{-1} \mathbf{Y} \right)
\end{equation}
where $\mathbf{Y}=\left[ \nu_c,\nu_r \right]$, and $\Sigma$ is the covariance matrix \footnote{When considering peaks theory, there are other relevant scalar variables which normally appear in the multi-variate Gaussian, such as the laplacian of $C_G$. However,they will not be important when we consider the high peak limit, and so we will not consider them further here.}.  

The PDF of the compaction, $P(C)$,  can be calculated by integrating the 2-variate Gaussian over the range of values of $C_G$ and $\zeta_r$ which give the specified value of $C$,
\begin{equation}
P(C) = \int\mathrm{d}C_G\mathcal{N}\left(C_G,\zeta_r(C,\nu_c,f,g)\right),
\end{equation}
where $\zeta_r(C,C_G,f,g)$ is expressed as a function of $C,\nu_c,f$ and $g$. In theory, this is the solution of a quartic solution, but as we will see, it is not necessary to calculate this in the high-peak limit.

To simplify the calculation, we can diagonalise the PDF by expressing it in terms of $\nu_c$ and $z$:
\begin{equation}
\mathcal{N}(\nu_c,\zeta_r) =\mathcal{N}\left(\nu_c\right)\mathcal{N}\left( z \right)= \frac{1}{\sqrt{2\pi}}\exp\left(-\frac{1}{2}\nu_c^2\right)\frac{1}{\sqrt{2\pi}}\exp \left( -\frac{1}{2}z^2\right),
\end{equation}
where $\mathcal{N}(z)$ can be expressed as
\begin{equation}
\mathcal{N}(z) = \mathcal{N}\left( \frac{\nu_r-\gamma_{cr}\nu_c}{\sqrt{1-\gamma_{cr}^2}} \right) = \frac{1}{\sqrt{2\pi(1-\gamma_{cr}^2)}}\exp\left( -\frac{\left( \nu_r - \gamma_{cr}\nu_c \right)^2}{2 \left(1-\gamma_{cr}^2\right)} \right).
\end{equation}
In the high-peak limit, $\nu_c\rightarrow \infty$,  then $\mathcal{N}(z)$ can be approximated as a Dirac-delta function:
\begin{equation}
\mathcal{N}(\nu_r) = \delta_D\left(  \nu_r - \gamma_{cr}\nu_c \right).
\end{equation}
This means that, when integrating over $\mathcal{N}(\nu_r)$, we can simply make the substition $\nu_r = \gamma_{cr}\nu_c$, or in terms of $C_G$ and $\zeta_r$ we can write,
\begin{equation}
\zeta_r = \gamma_{cr}\frac{\sigma_r}{\sigma_c}C_G.
\label{eqn:variableRelation}
\end{equation}

The net result of this is that, in the high-peak limit, we can write the linear compaction as
\begin{equation}
C_1 = C_G + \tilde{f} C_G^2 + \tilde{g} C_G^3,
\label{eqn:NGcomp}
\end{equation}
where we have introduced the notation $\tilde{f}=2f\gamma_{cr}\frac{\sigma_r}{\sigma_c}$ and $\tilde{g} = 3g\left(\gamma_{cr}\frac{\sigma_r}{\sigma_c}\right)^2$ for convenience. This takes a form similar to our original model of local-type non-Gaussianity, equation \eqref{eqn:localNG}.

%
%
\section{Calculating the abundance of primordial black holes in the high-peak limit}
\label{sec:NGabundance}
%
%

We will here consider the case of a delta-function peak in the power spectrum (corresponding to $\Delta\rightarrow 0$)
\begin{equation}
\mathcal{P}_\zeta(k) = \mathcal{A}k\delta_D\left( k - k_p \right).
\label{eqn:deltaPower}
\end{equation}
We note here that this is the power spectrum of the full, non-Gaussian, power spectrum. However, for the calculation of the PBH abundance, it will be much simpler to make use of the Gaussian component of the compaction. The variance of the Gaussian component of the compaction $\sigma_c^2$ to the variance of the linear, non-Gaussian, component of the compaction $\sigma_\mathrm{NG}^2$ as
\begin{equation}
\sigma_\mathrm{NG}^2 = \sigma_c^2 + 2\tilde{f}^2\sigma_c^4+6\tilde{g}\sigma_c^4+15\tilde{g}^2\sigma_c^6,
\label{eqn:sigmaNG}
\end{equation}
where $\sigma_\mathrm{NG}^2$ can be calculated from the curvature perturbation power spectrum as
\begin{equation}
\sigma_\mathrm{NG}^2= \frac{16}{81}\int\limits_0^k \frac{\mathrm{d}k}{k}(kr)^4  \tilde{W}^2(k,r)\mathcal{P}_{\zeta}.
\label{eqn:NGpower}
\end{equation}
This provides a simple method to determine the amplitude of the relevant moments of the power spectrum given in equation \eqref{eqn:moments}.

Assuming this form for the power spectrum allows us to make a number of simplifications to calculation and maintain analytic control of the calculation, whilst still giving an accurate calculation of the abundance:
\begin{enumerate}
\item Considering a narrow spectrum ensures that the assumption of the high-peak limit and spherical symmetry is valid (this is discussed in more detail in appendix \ref{sec:validity}).

\item Since there is only a single scale at which perturbations are large, we can neglect other scales. This means that we can neglect the criteria that perturbations have a particular scale, e.g.  we can neglect the $\delta_D\left( \mathrm{d}C/\mathrm{d}r \right)$ term in equation \eqref{eqn:peakConstraint} - and use traditional peaks theory \cite{Bardeen:1985tr}.

\item Additionally, we can conclude that, since peaks are spherically symmetric, we can apply the peak constraint to the Gaussian component of the compaction $C_G$ - assuming care is taken to integrate only over values corresponding to peaks and not troughs.

\item Whilst the delta-function power spectrum is unphysical \cite{Byrnes:2018txb},  the abundance of PBHs for lognormal peaks in the power spectrum with a width less than $\Delta\lesssim 0.3$ is well described by using the delta function \cite{Gow:2020bzo}.  Therefore, rather than considering a power spectrum of finite width, we can simply investigate a delta-function peak in the power spectrum without worrying about integrating over a range of scales at which PBHs form.
\end{enumerate}
We note that, whilst the calculation presented here could easily be extended to broad power spectrum (although one runs into the well-known problem that the variance $\sigma_c^2$ diverges for scale-invariant spectra), the consideration of such power spectra is left for future work. The results are expected to be qualitatively to previous work studying the effect of modal coupling in the context of broad power spectra and non-Gaussianity \cite{Young:2014oea,Tada:2015noa,Young:2015kda}.

For the narrow power spectrum in equation \eqref{eqn:deltaPower},  and setting the smoothing scale $r=2.74/k_p$, we obtain the following values for the required integrals of the power spectrum and correlation functions,
\begin{align}
\begin{split}
\sigma_c^2 &= (k_p r)^4  \tilde{W}^2(k_p,r)\mathcal{A}\simeq 2.01\mathcal{A},\\
\sigma_r^2 &=  \tilde{W}_s^2(k_p,r)\mathcal{A} \simeq 0.141\mathcal{A},\\
\sigma_{cr}^2 &= \frac{4}{9}(k_p r)^2 \tilde{W}(k_p,r)\tilde{W}_s(k_p,r) \mathcal{A}\simeq 2.00\times 10^{-1}\mathcal{A}.
\end{split}
\end{align}
Combining these gives us the factor appearing in equation \eqref{eqn:variableRelation},
\begin{align}
\gamma_{cr}\frac{\sigma_r}{\sigma_c}\simeq 9.95 \times 10^{-2},
\end{align}
gving us
\begin{align}
\tilde{f} \simeq 1.19 \times 10^{-1} f_\mathrm{NL}, \tilde{g}\simeq1.07\times 10^{-2}g_\mathrm{NL}.
\label{eqn:NGparams}
\end{align}
That the factor $\gamma_{cr}\frac{\sigma_r}{\sigma_c}$ is significantly less than unity implies that the impact of local-type non-Gaussianity on PBH abundance will be significantly less than has been calculated previously (such as in \cite{Byrnes:2012yx}), especially for higher order terms. The key reason for the difference is that the compaction is volume-averaged over the scale of the perturbation - and we are thus sensitive to the values of $\zeta$ at the edge of the perturbation, rather than the centre.

We note that, in this paper, we are neglecting changes to the profile shape of perturbations from non-Gaussianity (see \cite{Atal:2019erb} for a more detailed discussion of this effect). A changing profile shape would affect the threshold value for collapse as well as the mass scaling relationship (see equation \eqref{eqn:mass}).
It would also affect the scale at which the compaction peaks, and would therefore affect the smoothing scale, and the correlation factor. Using equation \eqref{eqn:NGparams} can therefore underestimate (overestimate) the effect of non-Gaussianity in the case that the non-Gaussianity parameters become large and positive (negative). A similar effect is discussed in more detail in reference \cite{Kitajima:2021fpq}.

In the high-peak limit, the number density of peaks of height in the range $C_G$ to $C_G+\mathrm{d}C_G$ is given by \cite{Bardeen:1985tr}
\begin{equation}
n(C_G) = \frac{1}{3^{3/2}(2\pi)^2}\left(\frac{\sigma_{1}}{\sigma_c}\right)^3 \left(\frac{C_G}{\sigma_c} \right)^3\exp\left( -\frac{C_G^2}{2\sigma_c^2} \right),
\label{eqn:numberDensity}
\end{equation}
where $\sigma_1/\sigma_c = k_p^2$ for a delta-function power spectrum. We note that, due to the symmetry of a Gaussian field, the number density of peaks of height $C_G$ will be equal to the number density of troughs of depth $-C_G$.

By considering that peaks in $C_G$ correspond to peaks (or troughs) in $C$, equation \eqref{eqn:numberDensity} will form the basis of our calculation going forwards.  The mass fraction of the universe which will collapse to form PBHs at the time of horizon entry is given by
\begin{equation}
\beta = \left( 2\pi \right)^{3/2}r^3\int \mathrm{d}C_G \frac{M_\mathrm{PBH}\left( C_G \right)}{M_H}n(C_G),
\label{eqn:beta}
\end{equation}
where the integral is performed over the range of values of $C_G$ which form PBHs.  Recalling that we will only consider the formation of PBHs from type I perturbations, which means that we will integrate over values of $C_G$ corresponding to values in the range $C_{1,th}<C-1<4/3$. $C_{1,th}$ is the threshold value of the linear component of the compaction, given from the compaction threshold $C_{th}$ by equation \eqref{eqn:C1ofC}, where for $C_{th}=0.50$, we obtain $C_{1,th}\simeq 0.67$. For the Gaussian case, the integration is therefore over the range $0.67<C_G<4/3$.

We note that, whilst equation \eqref{eqn:beta} is straightforwards to re-cast in terms of the compaction $C$, it is far simpler to perform the calculation using $C_G$ - which also allows us to differentiate between type I and type II perturbations.

\begin{figure*}
 \centering
  \includegraphics[width=0.45\textwidth]{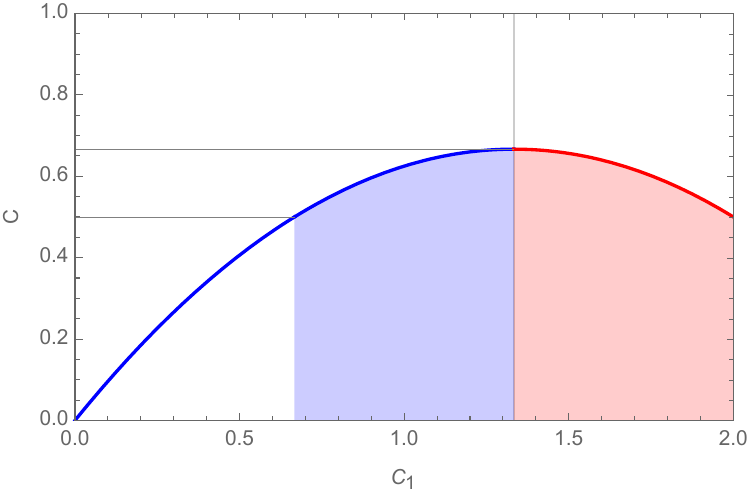}
\\
  \includegraphics[width=0.45\textwidth]{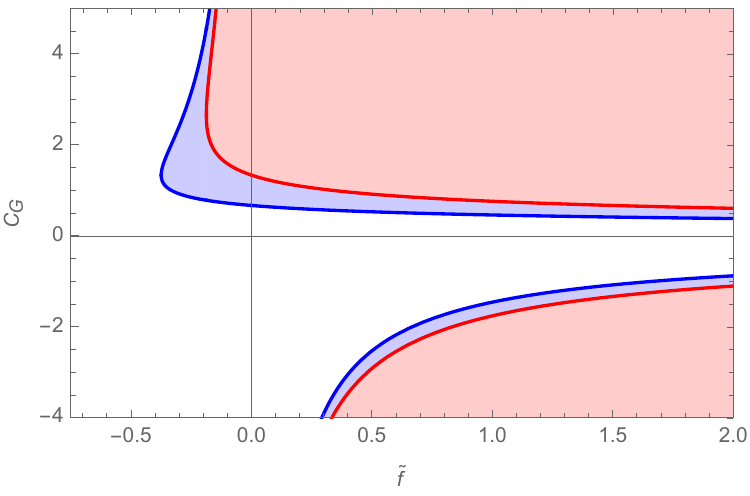}
  \includegraphics[width=0.45\textwidth]{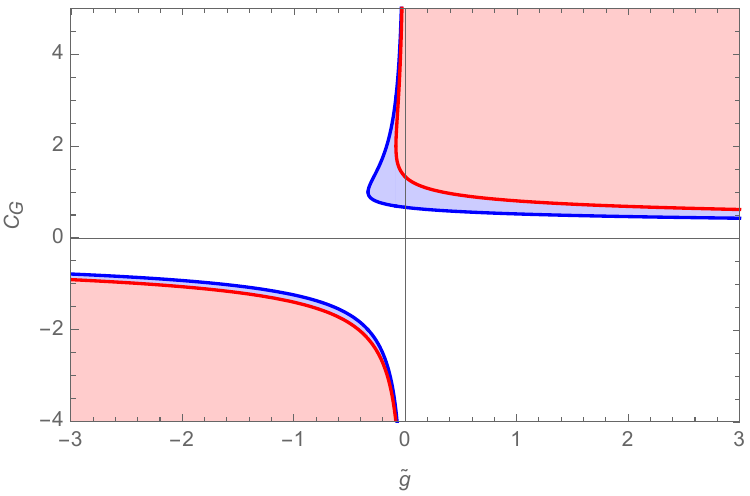}
  \caption{In all the plots, the blue region shows the values for PBH forming type I perturbations, and the red region shows the values for type II perturbations (which also form PBHs, although we neglect type II perturbations in the calculation).  
 \emph{Top plot}: the relation between the full non-linear compaction $C$ and the linear component $C_1$.  
\emph{Bottom-left plot}: the values of the linear, Gaussian component of the compaction $C_G$ which form PBHs, as a function of $\tilde{f}$ (and assuming $\tilde{g}=0$). 
\emph{Bottom-right plot}: similarly, the values of the linear, Gaussian component of the compaction $C_G$ which form PBHs, as a function  of $\tilde{g}$ (and assuming $\tilde{f}=0$).
 }
  \label{fig:intLimits}
  \vspace{1cm}
\end{figure*}

In the following sections, we will quantify the effect of local-type non-Gaussianity on PBH abundance by considering the quadratic and cubic terms independently. A consideration of combining the terms is again left for future study.

\subsection{Quadratic non-Gaussianity}

In this section, we will consider the effect of quadratic non-Gaussianity on the PBH abundance, setting $\tilde{g}=0$.  In this case, there are two solutions for expressing $C_G$ as a function of $C_1$:
\begin{equation}
C_G\left(C\right) =  C_\pm\left(C_1\right) = \frac{   -1 \pm \sqrt{1+4\tilde{f} C_1}    }{    2\tilde{f}    },
\end{equation}
where the two solutions will be identified using the subscript $\pm$, as used in middle equality above.

The limits on the integral in equation \eqref{eqn:beta} depend on the value of $\tilde{f}$:
\begin{itemize}

\item $\tilde{f}>-\frac{3}{16}$ (excluding $\tilde{f}=0$): PBHs form in the range $C_+(C_{1,th})<C_G<C_+(4/3)$ as well as in the range $C_-(4/3)<C_G<C_-(C_{1,th})$. 

\item $-\frac{1}{4C_{1,th}} < \tilde{f} \leq -\frac{3}{16}$: type II perturbations do not form in this regime, and we instead integrate over the range $C_+(C_{1,th})<C_G<C_-(C_{1,th})$.

\item $\tilde{f}\leq -3/8$: in this regime, there are no perturbations which form PBHs.

\end{itemize}
The integration limits are shown in the bottom left plot of figure \ref{fig:intLimits}. It is noteworthy that, except for large values $\tilde{f}\gg 1$, the abundance of PBHs is dominated by the value of $C_+(C_{t,th})$, and we could obtain an excellent approximation by simply integrating equation \eqref{eqn:beta} in the range $C_G>C_+(C_{t,th})$.

\begin{figure*}
 \centering
  \includegraphics[width=0.6\textwidth]{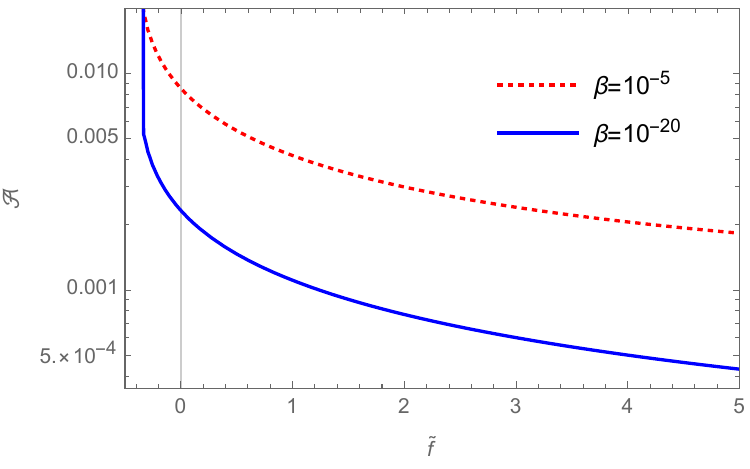}
  \caption{The amplitude of a delta-function power spectrum $\mathcal{A}$ required to produce a given initial abundance of PBHs $\beta$ as a function of the non-Gaussianity parameter $\tilde{f}$. }
  \label{fig:betaFnl}
\end{figure*}

This now allows us to relate the power spectrum $\mathcal{P}_\zeta$ to the variance of the Gaussian component of the compaction $\sigma_c^2$ using equations \eqref{eqn:sigmaNG} and \eqref{eqn:NGpower}, and then use equation \eqref{eqn:beta} to calculate the PBH abundance. Solving the integral numerically allows us to the amplitude of the power spectrum to the PBH abundance. Figure \ref{fig:betaFnl} shows the amplitude of the power spectrum required to produce PBHs in the abundance $\beta = 10^{-5},10^{-20}$ for varying values of $\tilde{f}$ (recalling $\tilde{f}\simeq 1.19 \times 10^{-1}f_\mathrm{NL}^\mathrm{local}$, for the specific case considered here).  

For negative $\tilde{f}$, the abundance of PBHs decreases rapidly, which means that a larger amplitude of the power spectrum $\mathcal{A}$ is required to produce the same number. For $\tilde{f}<-3/8$, there are no PBH forming perturbations, and thus the value of $\mathcal{A}$ diverges as we approach this limit. For positive $\tilde{f}$, the abundance of PBHs is significantly increased - and thus a smaller $\mathcal{A}$ is required to produce the same abundance. An alternative interpretation of the results is that, if one has a given bound on PBH abundance, for example, $\beta<10^{-20}$, then the constraints on the power spectrum become weaker (stronger) for negative (positive) $\tilde{f}$.

\subsection{Cubic non-Gaussianity}

We will now consider the effect of cubic non-Gaussianity on the PBH abundance, this time setting $\tilde{f}=0$. Since equation \eqref{eqn:NGcomp} is now cubic, there are 3 solutions for $C_G$ as a function of $C_1$, given by
\begin{align}
\begin{split}
C_G(C_1) = C_{a}(C_1) = \frac{  \left( \frac{2}{3} \right)^{1/3} \exp\left(i \theta_a\right)   }{   \lambda    }       -       \frac{   \exp\left( -i\theta_a \right) \lambda   }{   2^{2/3} 3^{1/3} \tilde{g}  },\\
\lambda = \left(    9\tilde{g}^2C_1 + \sqrt{   12\tilde{g}^3+81\tilde{g}^4C_1^2   }     \right)^{1/3},
\end{split}
\end{align}
where $\theta_a = \left[ \pi,  \pi/3, -\pi/3    \right]$ for $a = \left[ i,j,k\right]$.

As before, we find that the limits on the integral in equation \eqref{eqn:beta} depend on the value of $\tilde{g}$:
\begin{itemize}

\item $\tilde{g}\leq \tilde{g_c}$:  PBHs form in the range $C_i(4/3)<C_G<C_i(C_{1,th})$.

\item $-0.33 < \tilde{g} \leq -1/12$: PBHs form in the range $C_i(4/3)<C_G<C_i(C_{1,th})$ and $C_k(C_{1,th}<C_G<C_j(C_{1,th})$.

\item $-1/12 < \tilde{g}\leq 0$:  PBHs form in the range $C_i({4/3})<C_G<C_i(C_{1,th})$, $C_j(4/3)<C_G<C_j(C_{1,th})$ and $C_k(C_{1,th})<C_G<C_k(4/3)$.

\item $\tilde{g} > 0$:  PBHs form in the range $C_i(C_{1,th})<C_G<C_i(4/3)$.

\end{itemize}
where $\tilde{g}_c$ is given by
\begin{equation}
\tilde{g}_c = -\frac{4}{27C_{1.th}},
\end{equation}
and we obtain $\tilde{g}_c\simeq-0.33$ for $C_{1,th}\simeq 0.67$.
The integration limits are shown graphically in the bottom right plot of figure \ref{fig:intLimits}, and we again note that the PBH abundance is typically dominated by the solution $C_G(C_{1,th})$ with the smallest magnitude. For $\tilde{g}>0$, this is $C_i$, and for $\tilde{g}<0$, this is $C_k$.

\begin{figure*}
 \centering
  \includegraphics[width=0.6\textwidth]{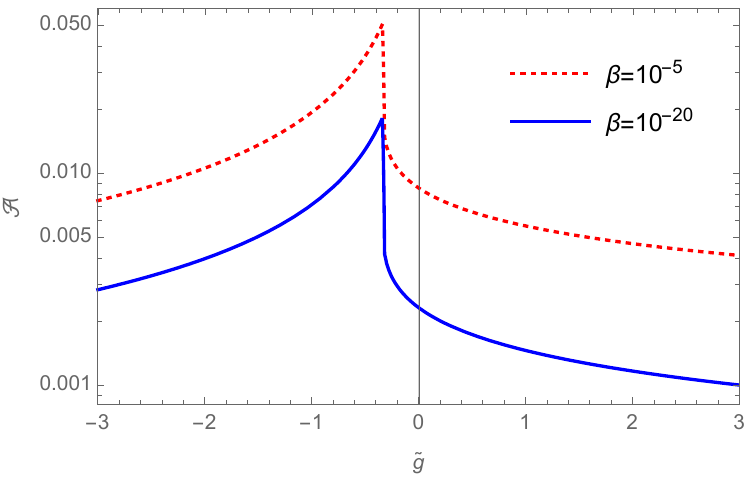}
  \caption{The amplitude of a delta-function power spectrum $\mathcal{A}$ required to produce a given initial abundance of PBHs $\beta$ as a function of the non-Gaussianity parameter $\tilde{g}$. }
  \label{fig:betaGnl}
\end{figure*}

The integral in equation \eqref{eqn:beta} can now be solved numerically to calculate the PBH abundance as a function of the power spectrum. Figure \ref{fig:betaGnl} shows the amplitude of the power spectrum required to produce PBHs in the abundance $\beta = 10^{-5},10^{-20}$ for varying values of $\tilde{g}$ (recalling now that $\tilde{g}\simeq 1.07 \times 10^{-2}g_\mathrm{NL}^\mathrm{local}$). 

We find that the abundance of PBHs is increased for positive $\tilde{g}$ (resulting in a smaller amplitude of the power spectrum required to produce the same abundance). For slightly negative $\tilde{g}$,  the abundance of PBHs decreases dramatically - resulting in a severe increase in the amplitude of the power spectrum required to produce the same abundance. This is due to the fact that, for $\tilde{g}<0$ there is a maximum amplitude of the compaction $C$ which can form from positive $C_G$, given by,
\begin{equation}
C_\mathrm{max} = \frac{ 1 }{  18\tilde{g}   }+\frac{  2 i  }{   3\sqrt{  3\tilde{g}  }   }.
\end{equation}
Switching from the regime where positive $C_G$ can form PBHs to the regime where they cannot results in the dramatic increase in the amplitude of the power spectrum $\mathcal{A}$ required to produce the same abundance of PBHs, as seen in figure \ref{fig:betaGnl}. For more negative values of $\tilde{g}$, we see that the abundance of PBHs starts to increase again, resulting in a smaller $\mathcal{A}$. 

For $\tilde{g}\rightarrow+\infty$ or $\tilde{g}\rightarrow-\infty$, the value of $\mathcal{A}$ asymptotes to the same value. This is because we can neglect the linear term and simply write $C_1 = \tilde{g}C_G^3$, which is invariant under the transformation $\tilde{g}\rightarrow-\tilde{g}$, $C_G\rightarrow-C_G$.

%
\section{Comparison to previous literature}
\label{sec:literature}
%

Qualitatively, the results are most similar to previous work by Byrnes et al \cite{Byrnes:2012yx} (which was followed up by a series of papers by Young and Byrnes \cite{Young:2013oia,Young:2014oea,Young:2015cyn,Young:2015kda}). The paper made use of a Press-Schechter-type approach and used the curvature perturbation as the formation criterion. Whilst this can be considered a valid approach for narrow power spectra (as is also considered here), it has since been argued that the density, and specifically the compaction should be used as the formation criterion, although there are many methods for performing the calculation \cite{Young:2014ana,Musco:2018rwt,Young:2019osy,Yoo:2018kvb,Yoo:2020dkz}.  Whilst the approaches used by Byrnes et al and this study are very different, the results are qualitatively very similar due to the similarity between equation \eqref{eqn:localNG} (which forms the basis for Byrnes et al) and equation \eqref{eqn:NGcomp} (which forms the basis for this study).  Quantitatively, we find that the effect of local-type non-Gaussianity can be an order of magnitude smaller than Byrnes et al found, which is due to the fact that the compaction is sensitive to the value of the curvature at the edge of a perturbation, rather than the peak value in the centre.

Riccardi et al \cite{Riccardi:2021rlf} makes use of the ``spiky enough" criteria to determine whether a perturbation in $\zeta$ will collapse to form a black hole. It achieves this by using equation \eqref{eqn:denCon} to relate $\zeta$ to the density contrast, and then effectively uses the density contrast as the formation criterion.  Using this approach, it is found that a positive $f_\mathrm{NL}^\mathrm{local}$ would actually suppress the formation of PBHs - in contradiction to the findings here and in previous papers.  

The contradiction is due to using the non-linear expression for the density contrast, equation \eqref{eqn:denCon}, and specifically, it is due to the $\exp\left( -2\zeta\right)$ term in the equation. When one includes an additional positive quadratic term to $\zeta$ (as in the local-type expansion, equation \eqref{eqn:localNG}) then, for the large, positive perturbations which form PBHs, this increases the value of $\zeta$ - which can therefore decrease the magnitude of the density contrast.  Since the addition of an $f_\mathrm{NL}^\mathrm{local}$ term decreases the amplitude of the density contrast, the conclusion is then that the abundance of PBHs will be decreased as well. However, the increased value of $\zeta$ also introduces a change to the horizon size whilst the perturbation is in the super-horizon regime. To illustrate this, let us consider the simple scenario of adding a constant value $\phi$ to the curvature perturbation, $\zeta\rightarrow \zeta+\phi$.  This decreases the value of the density contrast everywhere by a factor $\exp(-2\phi)$.  By applying the separate universe approach, we conclude that this should not affect the evolution of the universe, but instead simply introduces a time shift.  A given perturbation will then take longer to enter the horizon - and grows by an additional factor $\exp(2\phi)$ before horizon entry - exactly cancelling the effect of $\phi$.  This could be addressed, for example by smoothing over a specified areal radius,  where one obtains an expression proportional to the compaction, and would then find results compatible with those presented here.

Kitajima et al \cite{Kitajima:2021fpq} investigated the effect of local-type non-Gaussianity, finding that $f_\mathrm{NL}$ has a similar effect on the PBH abundance.  Although similar, their approach does differ in several key regards. Rather than using the compaction to determine the threshold value for PBH formation, they use the averaged value of the compaction, which has been argued to minimise the dependence on the profile shape \cite{Escriva:2019phb}.  The averaged compaction is related to the Laplacian of the curvature perturbation $-\Delta\zeta$ by assuming a typical profile shape for $\zeta$. Peaks theory is then used to calculate the number density of peaks in $-\Delta\zeta$ which form PBHs ($-\Delta\zeta$ is also used in the mass scaling relation instead of the compaction, as in equation \eqref{eqn:mass}).  For the monochromatic power spectra considered, this approach is entirely valid, but would run into complications when broad power spectra are considered since no smoothing is performed (see \cite{Young:2019osy} for more discussion).

%
\section{Conclusions}
\label{sec:conclusions}
%

The effect of local-type non-Gaussianity on the PBH abundance has been considered, which can also be applied to constraints on the primordial power spectrum derived from constraints on the PBH abundance. The effect of non-Gaussian corrections at second- and third-order were considered, with results broadly in line with previous work by Byrnes et al \cite{Byrnes:2012yx}. We have updated the calculation to account for recent developments in the field:
\begin{itemize}
\item the use of the compaction as the appropriate parameter to determine whether a perturbation will collapse to form a PBH;

\item the non-linear relationship between the compaction and the curvature perturbation;

\item the mass scaling relationship which relates the amplitude and scale of a perturbation to the eventual PBH mass;

\item and we have also included peaks theory in the calculation rather than a Press-Schechter-type approach.
\end{itemize}

We find that the effect of the non-Gaussianity parameters $f_\mathrm{NL}^\mathrm{local}$ and $g_\mathrm{NL}^\mathrm{local}$ is qualitatively similar to that found previously. Positive $f_\mathrm{NL}^\mathrm{local}$ increases the PBH abundance (tightening constraints), whilst negative $f_\mathrm{NL}^\mathrm{local}$ decreases the abundance (weaking constraints).  Positive $g_\mathrm{NL}^\mathrm{local}$ also increases the PBH abundance, whilst negative $g_\mathrm{NL}^\mathrm{local}$ can have varying effects. For small negative values, the PBH abundance is decreased significantly, but increases for large negative values.

However, by considering the compaction as the relevant parameter for PBH formation, we find that, quantitatively,  the non-Gaussianity parameters have a much weaker effect than found previously, and must be orders of magnitude larger to have the same effect. This is especially true when higher order terms are considered, and is due to the fact that the compaction is sensitive to the value of the curvature perturbation at the edge of the perturbation rather than the peak value (i.e.  the compaction includes the term $\zeta(r)$ rather than $\zeta(0)$).  

Previous papers have also studied the effect of modal coupling on the PBH abundance, which required artifical insertion of long-wavelength modes into the calculation (often utilising the peak-background split). Whilst not considered here,  if broad power spectra were to be considered, the formalism derived here automatically encodes the effect of such long-wavelength modes and the effect of modal coupling on the PBH abundance and mass function.

One result of considering such mocal-coupling is the formation of dark matter isocurvature modes if the PBH abundance and non-Gaussianity is not negligible. Previous papers found that this would place extremely strong constraints on the local-type non-Gaussianity parameters if even a small amount of PBHs exist \cite{Tada:2015noa,Young:2015kda}. Scch constraints would be made considerably weaker once the updated calculation presented here is accounted for - especially for the higher-order non-Gaussianity parameters.

We've assumed and justified spherically symmetry for the power spectrum considered here.  However,  as discussed in section \ref{sec:validity} we have shown that this assumption is not valid for broad power spectrum - revealing a problem with the calculation performed in many papers related to the assumption of the high-peak limit.  Whilst it may be expected that the expression for C in equation \eqref{eqn:quadratic} still holds at least approximately true for broad power spectra, this is an important consideration worthy of further study and will be the subject of future research.

\section*{Acknowledgements}
SY is supported by an MCSA postdoctoral fellowships, and would like to thank Subodh Patil for his helpful comments on a draft of this work. This project has received funding from the European Union’s Horizon 2020 research and innovation programme under the Marie Skłodowska-Curie grant agreement No 101029832.

\appendix

\section{The compaction}
\label{app:compaction}

The mass excess $\delta M$ contained within a sphere of areal radius $R$ can be calculated by integrating the density $\rho(\mathbf{x},t)$ over the volume of the sphere
\begin{equation}
\delta M = \int\mathrm{d}^3\mathbf{x}\rho(\mathbf{x},t) - \int\mathrm{d}^3\mathbf{x}\rho_b  = \int\mathrm{d}^3\mathbf{x}\rho_b\delta(\mathbf{x},t),
\label{eqn:MassIntegral}
\end{equation}
where we have substituted the density contrast $\delta=(\rho-\rho_b)/\rho_b$ in the second equality. The background density $\rho_b$ is assumed to have the critical density 
\begin{equation}
\rho_b = \frac{3H^2}{8\pi},
\end{equation}
where we are using natural units $c=G=1$ throughout the paper.

The density contrast $\delta(\mathbf{x},t)$ is related to the curvature perturbation $\zeta(\mathbf{x})$ as (see e.g. \cite{Musco:2018rwt})
\begin{equation}
\delta(\mathbf{x},t) = -\frac{2(1+\omega)}{5+3\omega}\left( \frac{1}{aH} \right)^2\mathrm{e}^{ -2\zeta(\mathbf{x})} \left( \nabla^2\zeta(\mathbf{x})+\frac{1}{2}(\bar{\nabla}\zeta(\mathbf{x}))^2 \right),
\label{eqn:denCon}
\end{equation}
where $\omega=1/3$ is the equation of state during radiation domination (which, for convenience, will be included purely as a numerical factor from here).  The time-dependance of the density contrast is encoded in the comoving horizon scale $(aH)^{-1}$.  Assuming spherical symmetry for the rare peaks which form PBHs, we can express the density contrast in terms of the radial coordinate $r$ as
\begin{equation}
\delta(r,t) = -\frac{4}{9}\left( \frac{1}{aH} \right)^2\mathrm{e}^{ -2\zeta(r)} \left( \zeta''(r)+\frac{2}{r}\zeta'(r)+\frac{1}{2}\zeta'(r)^2 \right),
\end{equation}
 where the prime denotes a derivative with respect to $r$.  We can express the areal radius as $R(r,t)=a(t)\mathrm{e}^{-\zeta(r)}r$, and, at the centre of spherically symmetric peaks, the expression for the compaction becomes
\begin{align}
\begin{split}
C(r) &= \frac{2}{R(r,t)}\rho_b(t)\int\limits_0^{R(r,t)} \mathrm{d}R \left[4\pi R(r,t)^2\right] \delta(r,t), \\
&=- \frac{2}{    a    \mathrm{e}^{\zeta(r)}r    }\frac{3H^2}{8\pi} \int\limits_0^r \mathrm{d}\left(  a   \mathrm{e}^{\zeta(r)}r  \right) \left[4 \pi \left(a   \mathrm{e}^{\zeta(r)}r\right)^2\right] \\
&\times\frac{4}{9}\left( \frac{1}{aH} \right)^2\mathrm{e}^{-2\zeta(r)} \left( \zeta''(r)+\frac{2}{r}\zeta'(r)+\frac{1}{2}(\zeta'(\mathbf{r}))^2 \right),  \\
&=- \frac{4}{3}r\zeta'(r)\left( 1+\frac{1}{2}r\zeta'(r) \right).
\end{split}
\end{align}

We will now describe an alternate derivation of the linear component of the compaction, $- \frac{4}{3}r\zeta'(r)$, allowing us to greatly simplify the calculation of the PBH abundance. To this end, we will first consider the linear expression for the density contrast in terms of $\zeta$
\begin{equation}
\delta_1(\mathbf{x}) = - \frac{4}{9}\left( \frac{1}{aH}  \right)^2\nabla^2\zeta(\mathbf{x}).
\end{equation}
Similar to the full expression,  integrating this expression over a sphere of areal radius $R=ar$ (neglecting the correction to the areal radius, $e^{\zeta}$, which gives a non-linear contribution) gives the linear calculation of the compaction
\begin{align}
\begin{split}
C_1(r) &=- \frac{2}{ar}\frac{3H^2}{8\pi} \int\limits_0^r\mathrm{d}\left(a r\right) 4\pi (a r)^2 \times \frac{4}{9}\left( \frac{1}{aH}  \right)^2\nabla^2\zeta(\mathbf{x}),\\
&= - \frac{4}{3}r\zeta'(r),
\label{eqn:linearCompaction}
\end{split}
\end{align}
which is the same expression as the linear component of the compaction when the full non-linear expression is used.  This allows us to write a simple expression for the full, non-linear, expression for the compaction in terms of the linear compaction $C_1$,
\begin{equation}
C(r) = C_1(r)-\frac{3}{8}C_1(r)^2,
\label{eqn:quadraticCompaction}
\end{equation}
which is equation \eqref{eqn:quadratic} in section \ref{sec:criteria}.

The expression for $C_1$ can also be expressed as a smoothing of the second derivative of $\zeta$ with a top-hat window function:
\begin{equation}
C_1(\mathbf{x},r)=\frac{4}{9}r^2 \int\mathrm{d}^3\mathbf{y} \nabla^2\zeta(\mathbf{y})W(\mathbf{x}-\mathbf{y},r),
\end{equation}
where the window function is given by
\begin{equation}
W(\mathbf{x},r) = \frac{3}{4\pi r^3}\Theta\left(  r - x   \right),
\end{equation}
where $\Theta(x)$ is the Heaviside step function.  Instead of then performing statistical calculations with the complicated non-linear expression (e.g. equation \eqref{eqn:denCon}), we can instead work with the (relatively) simple equation for $C_1$ given above, equation \ref{eqn:quadraticCompaction} (matchinig \ref{eqn:quadratic} in the text).

\section{Validity of the high-peak limit and the assumption of spherical symmetry}
\label{sec:validity}

In this section, we will discuss the validity of the assumption of the high-peak limit when calculating PBH abundance, as well as the related assumption of spherical symmetry.  In order that they don't dominate the density of the universe too quickly,  PBHs are necessarily rare at the time of their formation. In fact, the weakest constraint on the energy fraction of the universe contained in PBHs at the time of their formation is $\beta < \mathcal{O}(10^{-5})$, or for solar mass PBHs to make up the entirey of dark matter we have $\beta = \mathcal{O}(10^{-9})$.

What we can say then, is that the perturbations from which PBHs form are indeed rare. Since PBHs form at peaks in the compaction, we can safely conclude that peaks in the compaction are therefore rare, and that the high-peak limit is valid when dealing with the compaction. However, the derivation of equation \eqref{eqn:quadratic}, upon which a large number of papers are based, relies on the fact that perturbations in $\zeta$ are also spherically symmetric. Does the fact that PBH forming perturbations in the compaction are in the high-peak limit also imply that the corresponding perturbations in the curvature perturbation are in the high-peak limit?

We can investigate this by studying the correlation coefficient between the two different parameters, $C$ and $\zeta$. A correlation coefficient close to unity, $\gamma_{cr}\approx 1$,  signifies a strong correlation and would imply that a large value for $C$ at a specific location means that $\zeta$ also takes a large value at this location - which is to say that high peaks in $C$ would correspond to high peaks in $\zeta$.  Alternatively, should the correlation function be small, $\gamma_{cr}\ll 1$, then a large value for $C$ would imply nothing about $\zeta$. A large value for $C$ at a specific location could correspond to large, small, or even negative value for $\zeta$.

To examine the correlation function, we will consider a lognormal form form for the power spectrum which appears often in the literature,
\begin{equation}
\mathcal{P}_\zeta\left( k \right) = \frac{A}{\sqrt{2\pi}\Delta}\exp \left( - \frac{\ln\left(k/k_p\right)^2}{2\Delta^2} \right),
\label{eqn:lognormal}
\end{equation}
where $A$ determines the amplitude of the power spectrum, $k_p$ determines the location of a peak, and $w$ sets the width of the power spectrum. In the limit $\Delta\rightarrow 0$, one obtains the Dirac-delta form, and in the limit $\Delta\rightarrow\infty$ one obtains the scale-invariant power spectrum.

\begin{figure*}
 \centering
  \includegraphics[width=0.6\textwidth]{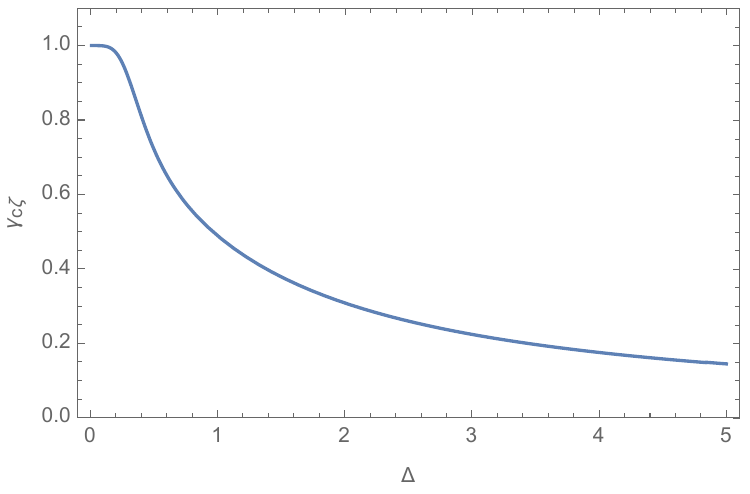}
  \caption{ The correlation coefficient $\gamma_{c\zeta}$ is plotted as function of the width of the power spectrum $\Delta$. We have used the power spectrum given in equation \eqref{eqn:lognormal}, with $k_p=1$ and smoothing scale $r=2.74/k_p$. }
  \label{fig:corrFunc}
\end{figure*}

The correlation coefficient $\gamma_{c\zeta}$ of the compaction $C$ and the curvature perturbation $\zeta$ is given by,
\begin{equation}
\gamma_{c\zeta} = \frac{\sigma_{c\zeta}}{\sigma_c\sigma_\zeta},
\end{equation}
which is a function of both the power spectrum $\mathcal{P}_\zeta$ and the smoothing scale $r$ - although is independent of the amplitude of the power spectrum. Figure \ref{fig:corrFunc} shows how $\gamma_{c\zeta}$ varies as a function of the power spectrum width $\Delta$, where we have used $r\simeq2.74/k_p$ \footnote{where we have chosen $r=2.74/k_p$ as this is the value at which $\sigma_c^2$ peaks for narrow power spectra $\Delta\rightarrow 0$, as well as corresponding to the scale of perturbations which form (although this is not true if the non-Gaussianity becomes large, which would affect the perturbation profiles, and consequently the scale at which the compaction peaks). We also note that, due to ringing in the window function, equation \eqref{eqn:FourierTH}, one can find a negative correlation coefficient when the smoothing scale $r$ is significantly larger than the scale at which the power spectrum peaks.}. For narrow power spectra,  $\gamma_{c\zeta}$ approaches unity, but is small for broad power spectra. In the limiting cases, we have
\begin{equation}
\lim_{\Delta\rightarrow 0}\left( \gamma_{c\zeta} \right)=1,\lim_{\Delta\rightarrow\infty}\left( \gamma_{c\zeta} \right)=0.
\end{equation}

The reason for this is clear: in the case of a narrow spectrum, there are only a small range of modes which can contribute to a given perturbation. Since all of these modes contribute to perturbations in both $C$ and $\zeta$, the two variables are strongly correlated. However, in the case of a broad power spectrum, there is a large range of modes which contribute to perturbations in $\zeta$. However, $C$ is only sensitive to a narrow range of scales - and is thus only dependant on a small subset of the modes comprising a perturbation in $\zeta$. In this case, the two variables are only weakly correlated.  

\begin{figure*}
 \centering
  \includegraphics[width=0.99\textwidth]{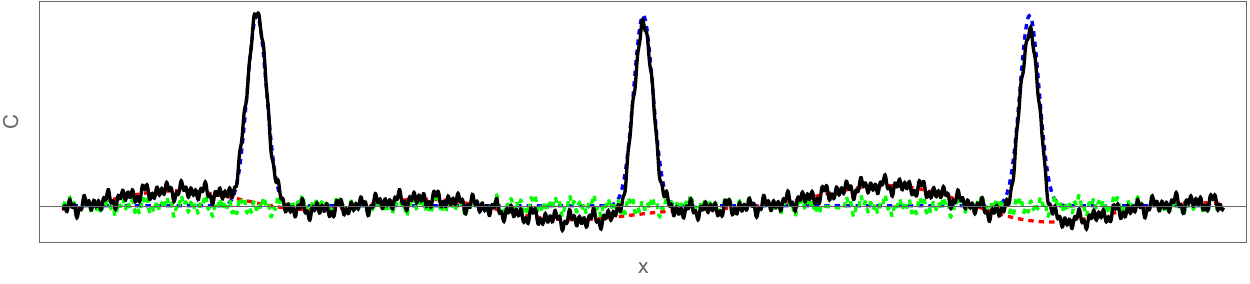}
  \includegraphics[width=0.99\textwidth]{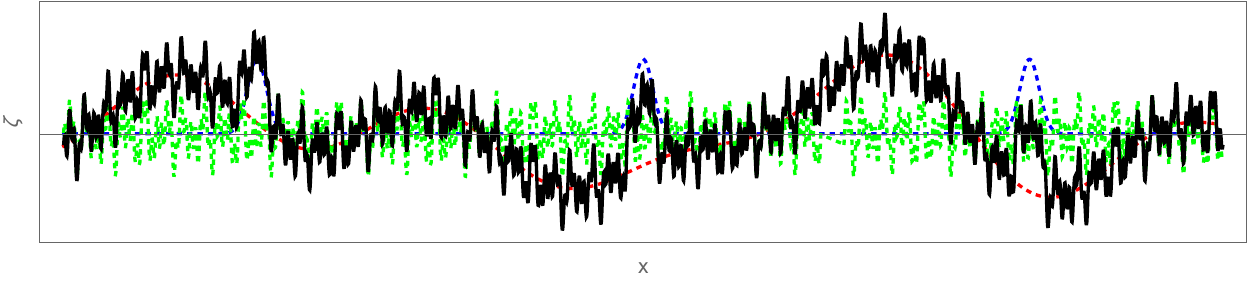}
  \caption{ A demonstration that high peaks in the compaction do not necessarily correspond to high peaks in the curvature. The top plot show a map of the compaction (with $x$ being some spatial coordinate), which contains randomly generated short (green) and long (red) wavelength modes, and to which 3 high peaks have been added on medium scales (blue) corresponding to the smoothing scale $R$. The total compaction is shown in black. The short and long wavelength modes are suppressed due to the nature of the compaction. The bottom plot shows the same perturbations,  but expressed in terms of the curvature perturbation.}
  \label{fig:demo}
\end{figure*}

This implies that, in the case of a broad power spctrum, high peaks in the $C$ do not correspond to high peaks in $\zeta$ - and that therefore the assumption of spherical symmetry of $\zeta$ is invalid.  This is demonstrated in figure \ref{fig:demo}. The top plot shows a schematic plot of a randomly generated map of the compaction function (black), starting from a broad power spectrum for the curvature perturbation. Three large amplitude peaks have been added by hand on the smoothing scale for the compaction.  Different scale modes have also been separated by dashed lines in different colours. The green line shows modes with a wavelength shorter than the smoothing scale, blue shows modes of approximately the smoothing scale, whilst red shows larger scale modes. The combined total is shown by the black line.  Due to the nature of the compaction, only modes close to the smoothing scale have a large effect\footnote{NB. We have not considered modes of very different scales, which could be large due to ringing in the window function.}. The bottom plot shows the same map, but expressed in terms of the curvature perturbation. 

The compaction $C$ is well correlated with modes in the $\zeta$ on the smoothing scale - peaks in the $C$ correspond to the blue component of $\zeta$, but shows little correlation with smaller or larger scale modes (in green and red). We can see that, whilst there are 3 easily identifiable high peaks in $C$, which are roughly symmetric, these do not correspond to high peaks in $\zeta$, and these are not symmetric.

We could therefore make the argument that equation \eqref{eqn:quadratic}, and everything that follows from it, is only valid for narrow power spectra, although such a statement is likely to be too strong. Whilst the derivation of equation \eqref{eqn:quadratic} does rely on the assumption of spherical symmetry, it is expected that an equivalent statement would still hold to be approximately true for non-symmetric perturbations. Nonetheless, in this paper we will take the conservative approach and consider only narrow power spectra, and leave the consideration of broad power spectra to future work.

\bibliographystyle{ieeetr}
\bibliography{bibfile}

\end{document}